\documentclass[intlimits,twoside,a4paper]{article}

\usepackage{amsmath,amssymb}
\usepackage{graphicx}
\usepackage[T2A]{fontenc}
\usepackage[cp1251]{inputenc}

\usepackage[eqsecnum]{cmpj2}
\usepackage{ulem}

\issue{2017}{20}{1}{13002}
\doinumber{10.5488/CMP.20.13002}
\title[On the decomposition theorem]%
{Helmholtz decomposition theorem and Blumenthal's extension by regularization}%

\author[D. Petrascheck, R. Folk]{D. Petrascheck, R. Folk}
\address{
Institute for Theoretical Physics, University Linz, Altenbergerstr. 69,
Linz, Austria
}

\authorcopyright{D. Petrascheck, R. Folk, 2017}
\date{Received December 1, 2016, in final form January 23, 2017}

\DeclareMathOperator{\diver}{div}
\DeclareMathOperator{\sgn}{sgn}

\begin{document}

\maketitle

\begin{abstract}
Helmholtz decomposition theorem for vector fields is usually presented with too strong restrictions on the fields and only for time independent fields. Blumenthal showed in 1905  that decomposition is  possible for any asymptotically weakly decreasing vector field. He used a regularization method in his proof which can be  extended to prove the theorem even for vector fields asymptotically increasing sublinearly.  Blumenthal's result  is then applied to the time-dependent fields of the dipole radiation and an artificial sublinearly increasing field.

\pacs 01.30.mm, 03.50.De, 41.20.Jb
\end{abstract}

\section{Introduction}

Regularization is nowadays a common method to modify the observable physical quantities  in order to avoid infinities and make them finite. Especially in the modern treatment of phase transitions by renormalization theory \cite{xxx} it is a tool in calculating, e.g., critical exponents. However, such regularization methods turned out to be also useful in university lectures on such classical fields as electrodynamics or hydrodynamics. Unfortunately, this method is rarely mentioned in this context. It is the aim of this paper to present
a classical example known as Helmholtz decomposition theorem and to show the power of regularization in this case.

According to the above mentioned theorem, one can divide a given vector field $\vec v(\vec x)$
into a sum of two vector fields $\vec v_l(\vec x)$ and $\vec v_t(\vec x)$ where $\vec v_l$ is irrotational (curl-free)
and $\vec v_t$ solenoidal (divergence-free), if the vector field  fulfills certain conditions on continuity and asymptotic decrease ($r\to \infty$).
Here, $\vec x$ is the position vector in three-dimensional space and $r=|\vec x|$ is its absolute value. Helmholtz calls these two components {\it integrals of first class} for which a velocity potential exists and {\it  integrals of second class}  for which this is not the case (\cite{helmholtz} first reference, p. 22).

Usually, these two integrals are constructed directly from the vector field by starting from the identity $\Delta \vec a = \vec \nabla \,\vec \nabla  \cdot \vec a - \vec \nabla \times (\vec \nabla \times \vec a)$ with  $\vec a = - \vec v(\vec x')/4 \pi|\vec x' - \vec x|$. Integrating $\Delta \vec a = \vec v(\vec x')\,\delta(\vec x - \vec x')$ over all space, one obtains for continuously differentiable vector field $\vec v$:
\begin{align} \label{v0}
\vec v(\vec x) &
= \underbrace{-\frac{1}{4 \pi} \int \rd^3 x '\, \vec \nabla \,  \vec \nabla \cdot\frac{\vec v(\vec x')}{|\vec x'-\vec x|}}_
{\textstyle \vec v_l(\vec x)}
+ \underbrace{\frac{1}{4 \pi} \int  \rd^3 x '\, \vec \nabla \times \left( \vec \nabla \times \frac{\vec v(\vec x')}{|\vec x'-\vec x|}\right)}_
{\textstyle \vec v_t(\vec x)}.
\end{align}
Or one calculates the two parts of the vector field from
 the respective potentials existing for them,
 \begin{equation}
 \vec v_l(\vec x) = - \vec \nabla \phi_{\text H}(\vec x) \, , \qquad \vec v_t(\vec x) = \vec \nabla \times \vec A_{\text H}(\vec x)\,  .
\end{equation}
These potentials are defined by the divergence and the rotation of the vector field
 \begin{align} \label{skalarpot1}
\phi_{\text H}(\vec x) &=   -\frac{1}{4\pi}\int \rd^3 x'\,\vec v(\vec x')\cdot
\vec \nabla' \frac{1}{|\vec x' - \vec x|} \, ,
\\ \label{vektorpot1}
\vec A_{\text H}(\vec x) &= \frac{1}{4\pi}\int \rd^3 x'\,\vec v(\vec x')\times
\vec \nabla' \frac{1}{|\vec x' - \vec x|}        \, .
\end{align}
As concerns validity, the uniqueness of decomposition and the existence of the respective potentials, one finds different conditions.

In fact, Helmholtz was largely anticipated by George Stokes (presented in 1849 and published in 1856 in \cite{stokes}, see p. 10, item 8), so it is also called Helmholtz-Stokes theorem \cite{johnson},
especially in hydrodynamics, where  the theorem is  of particular relevance.
There, the fluid fields of decomposition have physical properties of freedom of vorticity and incompressibility,
which for each field makes the analysis simpler \cite{lamb}.
In his discussion of the theorem,  Lamb \cite{lamb} states the conditions for divergency and vorticity of the vector field in infinity
in order to prove the theorem: they should be of the order of $1/r^{n}$ with $n>3$.

F\"oppl  introduced the decomposition theorem into German textbooks on electrodynamics \cite{abraham}.
In the first chapter he pre\-sents the appropriate tools of vector analysis since they were already used in hydrodynamics.
Regarding the theorem he assumed a finite extension of the sources and vortices and, therefore, assumed a behavior for the corresponding
vector field of the form $|\vec v| \sim 1/r^2$ for  $|\vec x|=r \to \infty$.
However, his proof permits less restrictive conditions, namely an asymptotic decay of the field only somewhat stronger than $1/r$.
The decomposition theorem can be found in one of these formulations in most textbooks or lecture notes on electrodynamics.

The main point made after presenting the theorem is in most cases the advantage of introducing a scalar and vector potentials.
It is applied in electrostatics and magnetostatics for cases where the extension of the sources is restricted to a finite region
(see for example \cite{miller}).
However, even in electro- and magnetostatics there exist configurations with slow decreasing fields.
The electric field of an infinite  straight wire, which bears an electric charge, decays as $\sim 1/\rho$, where $\rho$ is the distance to the wire.
If, on the other hand, the wire carries a current, then the magnetic field decays as $\sim 1/\rho$.
In both cases, a regularization is appropriate to get the potentials from finite integrals over the sources
without using symmetry arguments, which are not applicable in more complicated geometries.
A less restrictive formulation is found in \cite{griffith} (Appendix B as an interesting corollary)
stating that the field should go in infinity faster to zero than $1/r$.

Already in Aachen in 1905 professor Otto Blumenthal together with Sommerfeld,  proved \cite{blumenthal} that any vector field that goes to zero asymptotically
can be decomposed in a curl-free and a divergence-free part (weak version).
Blumental's formulation reads as follows (see \cite{blumenthal}, p. 236):

``Let $\vec v$ be a vector, which is, in addition to arbitrary many derivatives, everywhere finite and continuous
and vanishes at infinity with its derivatives; then one can always decompose this vector into two vectors,
a curl-free $\vec v_l$ and a divergence-free $\vec v_t$, such that
\begin{align} \tag{$^*$} \label{theorem1}
\vec v = \vec v_l (\vec x) + \vec v_t (\vec x) \, .
\end{align}
The vectors $\vec v_l$ and $\vec v_t$ diverge asymptotically weaker than $\ln r$.\\
In addition, one has  the following proposition for uniqueness:
$\vec v_l$ and $\vec v_t$ are unique up to an additive constant vector because of
the given properties.''  No further specification for the behavior of the vector field was given.

This formulation was taken over in its essential statements by Sommerfeld  in 1944 \cite{sommerfeld}.
He noted further that the fundamental theorem of vector analysis, as he called it,
was already proven by Stokes \cite{stokes} in 1849 and in a more complete form by Helmholtz paper of 1858.
 In a footnote he cites the paper of Blumenthal: {\it For a rigorous proof see: O. Blumenthal, Ueber die Zerlegung unendlicher Vektorfelder, (Math. Ann., 1905, \textbf{61}, 235). His only restriction is that $V$ and its first derivative vanish at infinity while no additional assumption is made how quickly they vanish. It turns out that the component fields $V_l$ and $V_t$ need not vanish themselves, they may even become in a restricted way infinite. In the following we shall make the somewhat vague assumption that $V$ vanishes ``sufficiently strongly'' at infinity.}

Later on it was shown  that the conditions of continuity and differentiability can be weakened \cite{butzer,bhatia}
and that the theorem can be applied to vector fields behaving according to a certain power law \cite{tran-cong}.
Based on Blumenthal's method of regularization of the Green function, Neudert and Wahl \cite{wahl}
among other things investigated the asymptotic behavior of a vector field $\vec v$ if its sources $\diver\vec v$ and vortices curl $\vec v$
fulfill some conditions including differentiability and asymptotic decay.

These developments remained to a large extent unnoticed in the physical literature  and
in mathematical physics (for an exception see \cite{gross}).
Thus, it seemed to be necessary to show the validity of the decomposition theorem for electromagnetic radiation fields that decay asymptotically like $1/r$. In fact there were several items to clarify for time dependent vector fields, especially the question of  retardation,
its connection to causality and the choice of gauge.

The paper is organized as follows: first we develop a systematic method of regularization,
then we reformulate the decomposition theorem including all potentials for such cases and finally we give two applications of the theorem.

\section{Regularization Method \label{met}}

The  regularization method, which is the basis of Blumenthal's proof, was not explicated in its generality and in its improvement  in order to be applicable to  vector fields, which decay  asymptotically with a specified power law (or even increase as we shall see below).
The idea is as follows: Since the property of the vector field cannot be changed in order to make the  involved integrals finite, one tries to change the weighting function $1/|\vec x' - \vec x|$ appearing in the solution for the two fields. Going back to the construction of these solutions, one used the Green function of the Poisson equation.

The solution $\phi_0(\vec x)$ of the Poisson equation
\begin{align} \label{poisson}
\Delta \phi_0 (\vec x)= -4\pi \rho(\vec x)
\end{align}
with the source density $\rho(\vec x)$ is found by introducing its Green function
\begin{align}\label{g0}
G_0(\vec x,\vec x')&= \frac{1}{|\vec x' - \vec x|} \, , \\
\phi_0(\vec x)&=   \int \rd^3 x'\,\rho(\vec x')\,G_0(\vec x,\vec x') \, .
\end{align}
If the solution exists in the whole domain of $\mathbb{R}^3$, the integral should be finite.
This is guaranteed by a sufficient decay of the integrand, either by a sufficient strong decay of the source density
and/or by a sufficient decrease of the Green function.

In his work on the Helmholtz decomposition theorem \cite{blumenthal}, Blumenthal presented a method to make this solution finite
(regularizing the solution) by changing  the Green function of the Poisson equation, without changing the Poisson equation
(which means without changing the source density). He  mentioned on p.~236 of \cite{blumenthal} the similarity of his method to
the ``convergence generating''  terms in the theorem of Mittag-Leffler.
Thus, one can prove the existence of the potential for cases where the source density is less strongly decreasing.  From this method it becomes clear how a systematic extension of the decomposition theorem  is possible.

Introduction of an arbitrary point $\vec x_0$ [apart from the condition that $\rho(\vec x_0)$ is finite at this point;
regularization point or regulator] and noting that
$G_0(\vec x,\vec x')=G_0(\vec x  - \vec x_0,\vec x'-\vec x_0)$,
we expand $G_0$ in a power series in $\vec x -\vec x_0$
\begin{align} \label{exp}
G_0(\vec x,\vec x')& = \frac{1}{|\vec x' - \vec x_0|} - (\vec x - \vec x_0)\cdot \vec \nabla' \frac{1}{|\vec x' - \vec x_0|} + \ldots
\, .
\end{align}
A stronger  decrease for large $|\vec x'|$ of the Green function is now reached by subtraction of the corresponding
expansion terms.
We get the following set of stronger decreasing Green functions
\begin{align}  \label{g1}
G_1(\vec x-\vec x_0,\vec x'-\vec x_0)&= G_0(\vec x,\vec x')- \frac{1}{|\vec x'  - \vec x_0|}\,,
\\  \label{g2}
G_2(\vec x-\vec x_0,\vec x'-\vec x_0)&
=G_1(\vec x-\vec x_0,\vec x'-\vec x_0)-\frac{(\vec x - \vec x_0) \cdot( \vec x'-\vec x_0)}{|\vec x'-\vec x_0|^3}
\, .
\end{align}
The asymptotic decrease of these modified Green functions is as $\sim 1/(r')^{1+i}$.
For $i \leqslant 2$, the subtracted terms do not change the source density
\begin{align} \label{deltagi}
\Delta G_i(\vec x -\vec x_0,\vec x'-\vec x_0) = - 4 \pi \delta(\vec x'- \vec x) \qquad
\text{for} \qquad 0 \leqslant i \leqslant 2 \, .
\end{align}
However, they make it possible to extend the range of the validity for which the existence of the potential (and the decomposition)
can be proven
\begin{align} \label{phierweitert}
\phi_i(\vec x)= \int \rd^3 x'\,\rho(\vec x')\,G_i(\vec x-\vec x_0,\vec x'-\vec x_0)
\qquad \text{and} \qquad
\Delta \phi_i(\vec x) = - 4 \pi \rho(\vec x) \qquad \text{for} \qquad i \leqslant 2
  \, .
\end{align}
The solutions $\phi_i(\vec x)$ differ only by a (divergence- and curl-free) solution of the Laplace equation,
i.e.,
$\phi_0(\vec x)$ differs from $\phi_1(\vec x)$ by a constant value and from $\phi_2(\vec x)$ by a linear function,
both depending on $\vec x_0$.
\par \smallskip
Trying to extend the range of validity even further, one may
subtract the next (third) term in the expansion  \eqref{exp} from $G_2$ and obtain
\begin{align}  \label{g3}
G_3(\vec x-\vec x_0,\vec x'-\vec x_0)&= G_2(\vec x-\vec x_0,\vec x'-\vec x_0)
-\frac{1}{2}\big ((\vec x - \vec x_0)\cdot \vec \nabla'\big )^2\,\frac{1}{|\vec x'-\vec x_0|} \,.
\end{align}
However, now $G_3$ fulfills the Poisson equation
\begin{align} \label{deltag3}
\Delta G_3(\vec x -\vec x_0,\vec x'-\vec x_0) = - 4 \pi \big [ \delta(\vec x'- \vec x)
- \delta(\vec x'-\vec x_0) \big ]
\end{align}
 from which it follows that $G_3$ leads to a solution of a modified Poisson equation
\begin{align}
\Delta \phi_3(\vec x) = - 4 \pi \big [ \rho(\vec x) - \rho(\vec x_0)\big]
  \, .
\end{align}
Thus, the method described here is not suitable for Green functions $G_i$ with $i > 2$.
This means (as we will see later) that  vector fields which increase linearly or even stronger
will not be decomposed by the regularization method described here.

Nevertheless, one should note that the Poisson equation can be solved even with $G_3$ if we
subtract the solution for the inhomogeneity $\rho(\vec x_0)$
\begin{align} \label{phi3}
\bar{\phi}_3(\vec x)=\int \rd^3 x'\,\rho(\vec x') G_3(\vec x - \vec x_0,\vec x'-\vec x_0) - 2\pi \frac{\rho(\vec x_0)}{3} \,|\vec x - \vec x_0|^2 .
\end{align}

The relation
\begin{align} \label{nabg2}
\vec \nabla G_{i+1}(\vec x-\vec x_0,\vec x'-\vec x_0) &= -\vec\nabla' G_i(\vec x-\vec x_0,\vec x'-\vec x_0),   \qquad i\geqslant 0,
\end{align}
can  be derived from  \eqref{g1}, \eqref{g2} and \eqref{g3}.
They  are used a few times, mainly to compute the vector fields $\vec v_l$ and $\vec v_t$ and to establish relations between them.

We would like to note that in higher order iterations of the regularization beyond $i=3$ for
the singularity of $G_i\sim |\vec x' - \vec x_0|^{-i}$ at $\vec x' = \vec x_0$, a convergence of the solution can only be reached
if the sources vanish sufficiently strongly at the regularization points.
In the following examples we will restrict ourselves to
a regularization for $i \leqslant 2$ at the point $\vec x_0=0$,
because the Green functions are simpler without loss of generality.
In this case, the scalar potential is fixed to  $\phi_i(\vec x=0) = 0$ for $i=1,2$.
We will keep this choice in the remaining part of the paper as far as possible.
\section{The extended fundamental theorem of vector analysis}

It has already been noted that today the formulation of the fundamental theorem rests in its form on the work of Blumenthal.
However, there are several reasons not to take the formulations of Blumenthal resp. Sommerfeld literally.
For instance, the uniqueness of the decomposition into the fields of the sources and vortices was only shown up to a constant vector.
We will formulate the conditions in such a form that a strict uniqueness of the decomposition is given.
Furthermore, in the proof that will be given below,  the potentials by which the decomposed fields are calculated
are part of the theorem (strong version).
It is common in electrodynamics to calculate the physical fields via the introduction of potentials.

Thus, we formulate the theorem in the following way:
\par \smallskip \noindent
{\sl Let $\vec v(\vec x)$ be piecewise continuous differentiable vector field,  then the decomposition}
\begin{equation}\label{vektorsatz}
\vec v(\vec x) = \vec v_l + \vec v_t = - \vec \nabla \phi_{\text H}(\vec x) + \vec \nabla \times \vec A_{\text H}(\vec x)
\end{equation}
{\sl reads}
\begin{align} \label{skalarpot}
\phi_{\text H}(\vec x) &=   -\frac{1}{4\pi}\int \rd^3 x'\, \vec v(\vec x')\cdot
\vec \nabla' G_i(\vec x-\vec x_0,\vec x'-\vec x_0) \, ,
\\ \label{vektorpot}
\vec A_{\text H}(\vec x) &= \frac{1}{4\pi}\int \rd^3 x'\, \vec v(\vec x')\times \vec \nabla'
G_i(\vec x-\vec x_0,\vec x'-\vec x_0) \, ,
\end{align}
{\sl where $i$ is taken for the asymptotic behavior
$\displaystyle \lim_{r \to \infty} v(r)\,r^{1-i+\epsilon} < \infty$,
 with $\epsilon >0$.
 This decomposition is unique for $i=0,1$ and unique apart for a constant vector field for $i=2$.}

\textit{Remarks}:
\begin{itemize}
\item Curl- and divergence-free fields $\vec v_h$ can be added to $\vec v_l$ if they are subtracted from $\vec v_t$
without affecting the boundary conditions of $\vec v$.
Such harmonic vector fields are suppressed if one explicitly demands that $\vec v_l$ and/or $\vec v_t$ should vanish asymptotically
and establish a strict uniqueness of the decomposition.
\item
Usually, the potentials $\phi_{\text H}(\vec x)$ and $\vec A_{\text H}(\vec x)$ are defined with the Green function $G_0$
\eqref{g0}.
If they are finite, then there is no need for $G_1$ \eqref{g1}.
However, if the vector field $\vec v$ decays asymptotically as $1/r$ or weaker,  one generally should
use the Green function $G_1$ as shown in  \eqref{skalarpot} and \eqref{vektorpot}
in order to avoid divergences in the potentials $\phi_{\text H}(\vec x)$ and $\vec A_{\text H}(\vec x)$.
\item As already mentioned in section \ref{met}, the potentials are for $i=1,2$ fixed  to the values $\phi_{\text H}(\vec x_0)=0$ and
$\vec  A_{\text H}(\vec x_0)= 0$ by the choice of the regularization point
$\vec x_0$, and for $i=1$ this
choice does not affect the vector fields $\vec v_l$ and $\vec v_t$, whereas for $i=2$ $\vec v_l$ and $\vec v_t$ they vanish at the
regularization point.
\item The vector potential $\vec A$ by its definition is purely transversal, $\vec \nabla \cdot \vec A_{\text H}=0$.
\item  In the special case of the theorem where $\vec v$ approaches zero at infinity weaker than any
power of $1/r$ (the case $\epsilon=1$), then $v_l$ and $v_t$ may diverge logarithmically although the sum of the two parts decays to zero
\cite{blumenthal}.
\item   We want to stress the point that the decomposition theorem holds for any vector field independent of the type of physical equations
that the vector field might fulfill.
On the other hand, if one thinks of the electric field or the magnetic field as examples of the theorem,
due to the Maxwell equations, these fields turn out to be connected although in relation to the decomposition theorem they are independent.
However, the  potentials for the decomposed parts can be identified with these fields.
\end{itemize}
Let us define the source density $\rho_{\text H}(\vec x)$ and the vortex density $\vec j_{\text H}(\vec x)$ as
\begin{align} \label{Hsource}
\rho_{\text H}(\vec x) = \frac{\vec \nabla \cdot \vec v(\vec x)}{4\pi}\,, \qquad
\vec j_{\text H}(\vec x) = \frac{\vec \nabla \times \vec v(\vec x)}{4\pi}\,,
\end{align}
then decomposition of the corresponding vector field in its irrotational (curl-free) and
solenoidal (diver\-gence-free) parts leads to the following result:
\begin{align}
\vec \nabla \cdot \vec v_l(\vec x)= 4\pi \rho_{\text H}(\vec x) \qquad &\text{and} \qquad \vec \nabla \times \vec v_l(\vec x)= 0 \, ,
\\
\vec \nabla \times \vec v_t (\vec x) = 4 \pi \vec j_{\text H}(\vec x) \qquad &\text{and} \qquad \vec \nabla \cdot \vec v_t(\vec x)=0 \, .
\end{align}
The potentials,  (\ref{skalarpot}) and (\ref{vektorpot}), can be rewritten by partial integration if the vector fields are everywhere continuously differentiable
\begin{align}
\phi_{\text H}(\vec x) = \int \rd^3 x' \,\rho_{\text H}(\vec x')G_i(\vec x-\vec x_0,\vec x'-\vec x_0), \qquad
\vec A_{\text H}(\vec x)  = \int \rd^3 x' \,\vec j_{\text H}(\vec x')G_i(\vec x-\vec x_0,\vec x'-\vec x_0).
\end{align}

The main advantage of the extended theorem lies in
the resulting systematic procedure of calculating the respective quantities.
This is done in the following way: One may start the integration with  $G_0$ for a {\bf finite} volume $V$.
If the integral does not converge in the limit $V \to \infty$, one should subtract the value of the  already calculated
quantity for the finite volume taken at the regularization point and then perform the limit $V \to \infty$, and so on.
Computing the scalar potential $\phi_{\text H}$ in \eqref{skalarpot} with $G_2$ needs no further integration
\begin{align}
\phi_{\text H}(\vec x,\vec x_0) &= \lim_{V\to \infty}\big[
\hat \phi_0(\vec x) - \hat \phi_0(\vec x_0) - (\vec x - \vec x_0)\cdot \vec \nabla_0  \hat \phi_0(\vec x_0) \big],
\end{align}
where $\hat \phi_0(\vec x)$ is the scalar potential calculated with $G_0$. It turns out that although the vector field is decaying at infinity of an order where a regularization seems to be necessary, the integrals might still converge and a further regularization is not necessary. The radiation field is such an example (see section~\ref{rad}).
\subsection{Sketch of the proof \label{phi21}}

We do not present explicit steps of the proof (for this see \cite{arxiv}),
but in order to show that the different integrals, which arise in \eqref{skalarpot} and \eqref{vektorpot},   exist and are finite,
we separate  the volume of integration into an inner volume of a sphere with radius  $R \gg r$ and the outer domain $r'  \geqslant R$.
Now, the large $r$ behavior of the corresponding $G_i$ is taken into account to prove the convergence.
We note that the singularities  at $\vec x$ and at zero do not lead to a diverging contribution to the integral,
as long as $i \leqslant 2$.
If the contribution of the outer domain to the potential vanishes,  then the existence of $\phi_{\text H}(\vec x)$ has been proved.

If the finiteness of the scalar potential \eqref{skalarpot} is  affirmed,
one gets  the field  $\vec v_l$ by calculating the gradient of $\phi_{\text H}$.
This field has the same sources as $\vec v$ ($\vec \nabla\cdot \vec v_l = \vec \nabla \cdot \vec v$)
and it is irrotational because the curl of a gradient field always vanishes.
Subsequently, one proceeds quite similarly for the vortex field $\vec A_{\text H}(\vec x)$.

Finally, we check that the sum $\vec v_l+\vec v_t=\vec v$ apart from a constant vector for $i=2$:
At first we switch in~\eqref{skalarpot} and \eqref{vektorpot} from $\vec \nabla'G_i$ to $-\vec \nabla G_{i+1}$ according to
\eqref{nabg2}. One obtains for the sum of $\vec v_l + \vec v_t$ using \eqref{deltagi} and \eqref{deltag3} for $x_0=0$
\begin{align}
\vec v(\vec x)+\vec v_t(\vec x)&=-\frac{1}{4\pi}\int \rd^3 x'\,\left [ \vec \nabla\vec \nabla\cdot \vec v(\vec x') - \vec \nabla \times \left(\vec \nabla \times \vec v(\vec x')\right)
\right ]G_{i+1}(\vec x,\vec x') = -\frac{1}{4\pi} \int \rd^3 x'\, \vec v(\vec x')\Delta G_{i+1}(\vec x,\vec x')
\nonumber\\
&=\vec v(\vec x) - \vec v(0)\delta_{i,2}\,.
\end{align}

\subsection{Comments on the uniqueness}

We have decomposed the vector field $\vec v$ in a source field $\vec v_l$ and a vortex field $\vec v_t$,
under the boundary condition that the total field $|\vec v|$ vanishes going to infinity.
In order to reach a uniqueness of the decomposition, we demand that
$|\vec v_l|$ and consequently also  $|\vec v_t|$  vanish going to infinity.
The respective differences of the longitudinal and transversal decomposition parts are divergence- and curl-free
and, hence, the harmonic solutions of the Laplace equation.
Due to the boundary condition in infinity, they should be zero and the differences of the vector fields are zero and the decomposition is unique.

An exception should be made in the case when $i=2$ is chosen.
Then, the difference in the vector fields could  be a linear harmonic function resulting in a uniqueness up to a linear term
(see again for more details in \cite{arxiv}).

\section{Application to time dependent fields and diverging fields}
\subsection{The radiation field \label{rad}}
When Blumenthal published his extension of the Helmholtz theorem, he pointed to the field of electromagnetic waves,
noting that it is of the $\mathrm{O}(1/r)$ and remarked:
{\it In consequence, for vector fields of this kind, the theorem in his present formulation would not be applicable.}%
\footnote{\it Auf derartige Vektoren w\"are also, der Satz in seiner bisherigen
Ausdehnung bereits nicht anwendbar.} Due to Blumenthal's proof, however, the theorem is applicable to such vector fields.

Usually, the theorem is not applied to time dependent problems in  textbooks on electrodynamics, whereas it is used in textbooks on hydrodynamics.
One reason might have been that the vector fields are solutions of Maxwell's equations which are relativistic contrary to the equations of classical fluid dynamics.

As late as the beginning of  the 21st century, this problem with the conventional formulation of Helm\-holtz theorem was taken up, without knowledge of Blumenthal's paper.
In literature one can find a discussion of the question whether the theorem can be applied to retarded fields. It was thought that this mathematical theorem could come into conflict with causality in the case of the propagation of time dependent vector field with finite velocity. The appearance of the quasistatic potentials has led to this discussion in the case of Coulomb gauge.
However it was recognized earlier \cite{brill} and confirmed later \cite{rohrlich}, that the physical quantities are causal and the decomposition is valid also for time dependent (retarded) fields. Rohrlich \cite{rohrlich} (see also  \cite{brill} and references therein) argued that the theorem can be applied to vector field of any time dependence, without referring to Blumenthal's paper and without mentioning the weaker decay of the radiation field.

However, the discussion went on considering the expressions of different options to choose the potentials for electromagnetic fields and it was shown by Jackson (see \cite{jack} and references therein) that quasistatic potentials can also be used. Nevertheless, the question was taken up again quite recently in a paper by Stewart \cite{stewart} with the title {\it ``Does the Helmholtz theorem of vector decomposition apply to the wave fields of electromagnetic radiation?''}. Since also in this paper Blumenthal's proof is not mentioned, the validity of Helmholtz decomposition is performed explicitly. This explicit calculation shows, on the other hand, that no regularization is necessary due to the appearance of $\re^{\ri k r}/r$ terms in the integrals.
Unfortunately, the author takes this property, which comes from the retardation, as an argument for nonconvergence of the integrals appearing in the Helmholtz decomposition for vector fields behaving as~$1/r$.

Radiation fields, which decay asymptotically as $1/r$, are rarely connected with the decomposition theorem.
If one starts with the assumption that the asymptotic behavior of the field should be stronger than $1/r$,
additional properties of the field are needed in order to prove the decomposition of the radiation fields \cite{stewart}.
Let us now show decomposition as an example of an oscillating point dipole. We also point to the differences in the meaning of different quantities such as sources and potentials within the theorem, (they are subscripted by ``H'') and those quantities appearing in Maxwell's equations and the potentials introduced to solve these equations.
Strictly speaking, the conditions of the theorem are not fulfilled if the vector field has singularities.
This also holds for the radiation fields considered below.
However, the integration over the sources in  \eqref{skalarpot} and \eqref{vektorpot} remains finite.

The periodically moving charge densities $\rho(\vec x,t)= \rho(\vec x)\,\re^{- \ri \omega t}$ of frequency $\omega$ emit a
radiation field of the same frequency.
For simplicity, we use the complex notation supposing that the physical quantities (charge density, potential, fields) are
always real parts of the corresponding complex quantities.
The radiation fields factorize in the same way as the sources $\vec v(\vec x,t)=\vec v(x) \,\re^{-\ri \omega t}$,
where in $\vec v(\vec x)$,  the dependence on the frequency $\omega$ resp. wave number $k=\omega/c$ has been suppressed.

The electric radiation field $\vec E(\vec x)$
of an oscillating point dipole $\vec p(t)=\vec p \, \re^{- \ri \omega t} $  reads in Gaussian units \{see \cite{petra},  (8.4.5) and (8.4.6), p.~294. The time dependence is in this case contained in the Fourier factor~$\re^{-\ri\omega t}$\}
\begin{align} \label{g5.41}
\vec v_E(\vec x) \equiv \vec E(\vec x) = \frac{\re^{\ri k r}}{r}\left \{ k^2  \vec e_r  \times (\vec p  \times \vec e_r)
  + \frac{1}{r^2}(1  - \ri kr )\Big [ 3 (\vec p  \cdot \vec e_r) \vec e_r  -\vec p \Big ]\right\} .\!
\end{align}
 $\vec e_r = \vec x /r$ is the unit vector in the direction of $\vec x$, and
$\vec v_E(\vec x)$ is the spatial part of the electric field.
We add the magnetic radiation field
\begin{align}
\vec v_B(\vec x)\equiv\vec B(\vec x)=k^2 \frac{\re^{\ri k r}}{r}\left (1-\frac{1}{\ri kr}\right)(\vec e_r  \times \vec p) \, .
\end{align}
Both fields are a solution of Maxwell's equations and fulfill
\begin{align} \label{maxw}
\vec \nabla \cdot \vec E(\vec x) =4\pi \rho(\vec x) \, , \qquad  \vec E(\vec x) =\frac{\ri}{k}\Big[\vec \nabla \times \vec B(\vec x) - \frac{4\pi}{c}\vec j(\vec x)\Big] \, ,  \qquad \vec \nabla\cdot\vec B (\vec x)=0 \, , \qquad \vec \nabla\times\vec E(\vec x)= \ri k\vec B(\vec x),
\end{align}
 the source and vortex density should be [see  \eqref{Hsource}]
\begin{align} \label{esource}
\rho_{E\text{H}} (\vec x) &
\widehat{=}\,\rho_p(\vec x) \, ,
 \qquad \rho_p(\vec x) = - \vec p  \cdot \vec \nabla \delta(\vec x ) \, , \qquad \vec j_{E\text{H}}(\vec x) = \frac{\ri k}{4\pi}\vec B(\vec x) \, ,
 \qquad \vec j_p(\vec x)=-\ri ck\vec p \delta(\vec x) \, .
\end{align}
$\rho_p(\vec x)$ is a localized charge  of the static dipole and $\vec j_p(\vec x)$ is the current density of the local oscillating dipole.
The Helmholtz vortex density is extended in the whole domain decreasing for  $r \to \infty$  with $1/r$.
 It can be identified with the spatial part of the magnetic radiation field $\vec B$ \{see \cite{petra},  (8.4.5) and (8.4.6), p.~294\} apart from a factor, as expected from Faraday's law of induction.
The wave number dependence in different quantities is caused by retardation.
Surprisingly, in $\rho_{E\text{H}}(\vec x)$, contrary to  $\vec j_{E\text{H}}(\vec x)$, it drops out.
This asymmetry has already been discussed by Brill and Goodman \cite{brill}.
Due to the absence of retardation in $\rho_{E\text{H}}$, the scalar potential is quasistatic
\begin{align}
\phi_{E\text{H}}(\vec x) &= -\frac{1}{4\pi}\int \rd^3 x'\,\vec v_E(\vec x')\cdot
\vec \nabla' \frac{1}{|\vec x' - \vec x|} =\int \rd^3 x'\,
\frac{\rho_{E\text{H}}(\vec x')}{|\vec x' - \vec x|}
 = \frac{\vec p  \cdot \vec e_r}{r^2}=\phi^{\text{qstat}}(\vec x)=\phi_{\text C}(\vec x) \, .
\end{align}
Multiplying by the factor $\re^{-\ri \omega t}$, one obtains the quasistatic (acausal) dipole potential $\phi_{\text C}(\vec x, t)$
as it is known using the Coulomb gauge.
Hence, it is clear that  the longitudinally decomposed vector field $\vec v_l$ is the quasistatic electric field of a point dipole
\begin{align}
\vec v_{El}(\vec x) & = - \vec \nabla \phi_{E\text{H}}(\vec x) = \big [ -\vec p  + 3 (\vec p  \cdot \vec e_r) \vec e_r \big ] \frac{1}{r^3}= \vec E^{\text{qstat}}(\vec x)
\end{align}
and does not contribute in the radiation zone to the electric field, which is  purely transversal.
The decomposition is finally shown by calculating the transversal part  $\vec v_{Et}= \vec \nabla \times \vec A_{E\text{H}}$ according to
the theorem of the vector potential
\begin{align} \label{ahele}
\vec A_{E\text{H}}(\vec x)&= \frac{1}{4\pi}\int \rd^3 x' \,\vec v_E(\vec x')\times \vec \nabla'\frac{1}{|\vec x'  - \vec x|}= \frac{\ri}{4\pi k}\int \rd^3 x' \,\left[\vec B(\vec x')\times\vec \nabla'- \frac{4\pi}{c}\vec j(\vec x')\right]\times \vec\nabla' \frac{1}{|\vec x'  - \vec x|}\nonumber\\
& = \frac{\ri}{k}\vec B(\vec x) + \frac{\vec e_r}{r^2} \times \vec p =  \frac{\ri}{k}\big[\vec B(\vec x) -  \vec B^{\text{qstat}}(\vec x) \big]  \, ,
\end{align}
where $\vec B^{\text{qstat}}(\vec x) $ is the quasistatic magnetic field of a point dipole.  We have again used Maxwell's equations~\eqref{maxw}.

In electrodynamics, one never defines a vector potential for the electric field, but it is known
from the Amp\`ere-Maxwell-equation that the electric field outside the sources
can be calculated via the curl of~$\vec B$. However, this is just the way we can calculate the transverse vector field
\begin{align}
\vec v_{Et}(\vec x) & = \vec\nabla\times\vec A_{E\text{H}}(\vec x)= \vec E(\vec x) - \vec E^{\text{qstat}}(\vec x) = \vec E(\vec x) - \vec v_{El}(\vec x) .\!
\end{align}
The causal character of the total electric radiation field $\vec v_E(\vec x)$ is restored \cite{rohrlich,jack}. This way of calculation is quite general and it is not only restricted to point sources.

The same decomposition may be done for the magnetic radiation field $\vec B(\vec x)$, which, however, is trivial since the field is only transversal [see \eqref{maxw}].
The Helmholtz vortex density of the magnetic field is presented by
 \eqref{esource}
\begin{align}
\vec j_{B\text{H}}(\vec x) = \frac{1}{4\pi}\vec \nabla \times \vec B(\vec x)=\frac{1}{c}\vec j(\vec x)-\frac{\ri k}{4\pi} \vec E(\vec x) \, ,
\end{align}
which is the total electric current (including the displacement current).

The Helmholtz vector potential fulfills  $\vec \nabla \cdot \vec A_{B\text{H}} = 0$ and $\vec \nabla \times A_{B\text{H}}=\vec B$, the same conditions as for the vector potential $\vec A_{\text C}$ in the Coulomb gauge. We indeed obtain for this example
$\vec A_{B\text{H}}(\vec x) = \vec A_{\text C}(\vec x)$ \cite{petra,arxiv}.
\begin{align} \label{vectorHC}
\vec A_{B\text{H}}(\vec x) &=\frac{1}{4\pi}\int \rd^3 x' \,\vec v_B(\vec x')\times \vec \nabla'\frac{1}{|\vec x'  - \vec x|}
=\frac{1}{\ri k}\big[\vec E (\vec x)-\vec E^{\text{qstat}} (\vec x) \big].
\end{align}

Thus, all the fields, the vector potential $\vec A_{B\text{H}}(\vec x)$, the vortex field
$\vec B(\vec x)= \vec \nabla \times \vec A_{B\text{H}}(\vec x)$
and the vortex density $\vec j_{B\text{H}}(\vec x)=\vec \nabla \times \vec B(\vec x)/4 \pi$
decay asymptotically as $1/r$. This is a consequence of retardation. In the Helmholtz vector potentials for both radiation fields,
one explicitly sees that the corresponding  quasistatic parts without retardation are subtracted.

One may be surprised that all calculations for the radiation field could be performed without regularization as expected according to the order of the decay of the vector field. Anyway, the integrals converge in an explicit calculation \cite {stewart}.
This might happen in other cases too ($G_0$ instead of $G_1$ etc.). One reason lies in the symmetries of the sources and circulations.
For instance,  fields like $\vec v(\vec x)= \vec p/r$ need no regularization. On the contrary, for a vector field like
$\vec v(\vec x)= \vec e_r/r$, the regularization term is necessary to reach convergence,
but the regularization point $\vec x_0$ should be different from zero. In such a case, we get for the potential
$ \phi(\vec x)=\ln r_0 - \ln r$.

\subsection{Finite or diverging fields in infinity}
In order to  demonstrate the extended theorem, we present two mathematical examples. Both of them have the same vectorial structure as the radiation field.
Let us take first the vector field
\begin{align} \label{constfield}
\vec v(\vec x) &= \vec e_r \times (\vec a \times \vec e_r),
\end{align}
where $\vec a$ is  a constant vector. $\vec v$ is finite but nonzero in the limit $r \to \infty$.
It seems to be more convenient to firstly determine sources and vortices
and then to calculate the fields belonging to these
\begin{align}
\rho_{\text H} (\vec x)
=-\frac{1}{4\pi}\frac{2}{r}(\vec a \cdot \vec e_r) \, , \qquad
\vec j_{\text H}(\vec x)
=\frac{1}{4\pi}\frac{1}{r}(\vec e_r \times \vec a) \, .
\end{align}
We would like to note that the example includes some subtle items:
(1) Since the vector field is not continuous at the origin, we should take $\vec x_0$ different from zero as regularization point.
(2) Due to the vectorial character of the field, the singularity at zero in the source and vortices is approached differently.
This is no obstacle for applying the theorem.
Then, we get $\phi_{\text H}$ from  \eqref{skalarpot} as follows:
\begin{align} \label{bsp.phi}
\phi_{\text H}(\vec x,\vec x_0) &= \int \rd^3 x'\,\rho_{\text H} (\vec x')  G_2(\vec x -\vec x_0,\vec x'-\vec x_0) =  - \frac{2}{3}\left \{(\vec a  \cdot \vec x)  (\ln r_0  - \ln r )
- (\vec a  \cdot \vec e_{r_0})\big [r_0
 -  (\vec x  \cdot \vec e_{r_0})\big ] \right \}.
\end{align}
To calculate the integrals, it is useful to introduce spherical coordinates.
The analogous calculation for the vector potential \eqref{vektorpot} yields
\begin{align}
\vec A_{\text H}(\vec x,\vec x_0) &
= \frac{1}{3}\left \{(\vec x  \times \vec a) ( \ln r_0  - \ln r)
  - (\vec e_{r_0}  \times \vec a) \left [ r_0
- (\vec x  \cdot \vec e_{r_0})  \right ] \right \} .
\end{align}
In   the last step, i.e., the calculation of the decomposed vector fields,
we get
\begin{align}
\vec v_l(\vec x,\vec x_0) = - \vec \nabla \phi_{\text H}(\vec x,\vec x_0 )\equiv \vec v_l (\vec x) -\vec v_l (\vec x_0)
\qquad &\text{with} \qquad
\vec v_l(\vec x) = \frac{2}{3}\big [-\ln r \vec a + \vec v(\vec x)\big ],
\\
\vec v_t(\vec x,\vec x_0) = \vec \nabla \times \vec A_{\text H}(\vec x,\vec x_0)\equiv \vec v_t (\vec x) -\vec v_t (\vec x_0)
\qquad &\text{with} \qquad
\vec v_t(\vec x)=\frac{1}{3}\big [ 2 \ln r\,\vec a  + \vec v(\vec x)\big ].
\end{align}
Thus, we have demonstrated that the vector field can be decomposed in
its irrotational and solenoidal components, both diverging logarithmically.
However, these terms cancel in the sum and it is indeed $\vec v_l(\vec x) +\vec v_t(\vec x) =\vec v(x)$.
A similar calculation can be performed for the example of a sublinearly diverging vector field
\begin{align} \label{diverginfield}
\vec v(\vec x) = \sqrt{r}\,\vec e_r \times (\vec a \times \vec e_r)
\, .
\end{align}
Now, the vector field is continuous at the origin and, therefore, one is allowed to choose the regularization point $\vec x_0=0$.  The decomposition reads
\begin{align}
\vec v_l(\vec x)
 =  - \frac{4\sqrt{r} }{7}\big [2\vec a + (\vec a\cdot \vec e_r) \vec e_r\big], \qquad
\vec v_t(\vec x)
= \frac{4\sqrt{r} }{7}\big [ 2\vec a+ \ (\vec a\cdot \vec e_r)\vec e_r\big ]+ \vec v(\vec x)\, .
\end{align}
\section{Conclusion}

We have presented the fundamental theorem of vector analysis (Helmholtz decomposition theorem)
for vector fields decaying weakly  and extended it to even sublinearly diverging vector fields by a systematic regularization procedure.
Contrary to the original proof \cite{blumenthal}, we can  distinguish between different cases.
Note, however, that not only the decay of the vector field is important
for introducing a regularization but also its symmetry. So, it might be the case that due to symmetry reasons, a lower level of regularization can be used in the decomposition as might have been expected just looking at the order of the decay of the vector field.

Thus, considering the validity of Helmholtz decomposition theorem, there is no doubt that the
theorem can be quite generally applied to electromagnetic fields either static or dynamic.
Because of the relevance of this extension of the Helmholtz decomposition theorem for textbooks on
electrodynamics and on mathematical physics, a pedagogical version has been given by one of the authors \cite{petranew}.

There are physical examples in electro- and magnetostatics with sources which extend to infinity and strength does not decay to zero there,
like a charged straight wire.
Usually, the fields of highly symmetric examples can be calculated in reduced geometry (e.g., in two dimensions).
We only mention that the vector field  for a charged $xy$-half plane [$\rho_{\text H}(\vec x) = \sigma \delta(z) \theta(x)$]
\begin{align}  \label{e}
\vec { v}(\vec x) &= \frac{ \sigma }{4\pi}\left [ \vec e_x \, \ln \left(x^2  + z^2\right)
+ \vec e_z  \left( \pi \sgn z  + 2  \arctan \frac{x}{z}\right)
\right ]
\end{align}
diverges logarithmically in the $x$-direction,
but can be calculated using the formalism of the Helmholtz decomposition theorem.


\section*{Acknowledgements}
Since we know the broad interest of Yurij also in historical and pedagogical topics in physics it is a great pleasure for us
to have the opportunity to present him this paper to the 60th birthday.
One of us (R.F.) thanks for a longstanding fruitful cooperation.


\newpage
\ukrainianpart

\title{Теорема про розвинення  Гельмгольца і регуляризаційне розширення Блюменталя}%

\author{Д. Петрашек, Р. Фольк}
\address{
Інститут теоретичної фізики, Університет м. Лінц, м. Лінц, Австрія
}

\makeukrtitle

\begin{abstract}
Теорема про розвинення Гельмгольца для векторного поля зазвичай представляється з сильними обмеженнями на поле і лише
для незалежних від часу полів. У 1905 р. Блюменталь показав, що розвинення є можливим для любого асимптотично слабоспадного векторного
поля. Він використав у доведенні регуляризаційний метод, який можна було розширити для доведення теореми для векторних полів, що є
асимптотично сублінійно висхідними. Результат Блюменталя застосовано до часовозалежних полів  дипольного випромінення і
штучного сублінійно висхідного поля.

\keywords  теорема Гельмгольца, векторне поле, електромагнітне випромінення
\end{abstract}
\end{document}